\documentstyle[aaspp4,epsf,12pt]{article} 

\newcommand{\xte}{{\sl RXTE}}
\newcommand{\psec}{s$^{-1}$}
\newcommand{\fouru}{4U~1702$-$429}

\begin{document}
\title{Observation of Kilohertz Quasiperiodic Oscillations from
      the Atoll Source \fouru\ by \xte}

\author{Craig B. Markwardt\altaffilmark{1,2}, Tod E. Strohmayer\altaffilmark{1}, Jean H. Swank\altaffilmark{1}}
\altaffiltext{1}{NASA/Goddard Space Flight Center, Code 662, Greenbelt, MD 20771; craigm@lheamail.gsfc.nasa.gov (CBM); stroh@lheamail.gsfc.nasa.gov (TES); swank@pcasun1.gsfc.nasa.gov (JHS)}
\altaffiltext{2}{National Research Council Resident Associate}

\begin{abstract}

We present results of {\it Rossi X-Ray Timing Explorer} (\xte)
observations of the atoll source \fouru\ in the middle of its
luminosity range.  Kilohertz-range quasiperiodic oscillations (QPOs)
were observed first as a narrow (FWHM $\sim$7~Hz) peak near 900~Hz,
and later as a pair consisting of a narrow peak in the range
625--825~Hz and a faint broad (FWHM 91~Hz) peak.  When the two peaks
appeared simultaneously the separation was $333\pm 5$~Hz.  Six type~I
thermonuclear bursts were detected, of which five exhibited almost
coherent oscillations near 330~Hz, which makes \fouru\ only the second
source to show burst oscillations very close to the kilohertz QPO
separation frequency.  The energy spectrum and color-color diagram
indicate that the source executed variations in the range between the
``island'' and ``lower banana'' atoll states.  In addition to the
kilohertz variability, oscillations at $\sim$10, $\sim$35, and 80~Hz
were also detected at various times, superimposed on a red noise
continuum.  The centroid of the $\sim$35~Hz QPO tracks the frequency
of the kilohertz oscillation when they were both present.  A
Lense-Thirring gravitomagnetic precession interpretation appears more
plausible in this case, compared to other atoll sources with low
frequency QPOs.

\end{abstract}

\section{Introduction}

Since the discovery of kilohertz (400--1200 Hz) quasiperiodic
oscillations (QPOs) from 4U~1728--34 (Strohmayer et al 1996) and Sco
X-1 (van der Klis et al 1996) with the {\it Rossi X-Ray Timing
Explorer} (\xte), there has been a concerted effort to characterize
the phenomenon by examining a variety of low mass X-ray binary
systems.  To date nearly twenty such sources have been discovered (van
der Klis 1998; Strohmayer, Swank, \& Zhang 1998), and a general
pattern has emerged.  In nearly all of the sources two kilohertz QPO
peaks are seen.  Both QPO frequencies increase and decrease as the
source intensity and spectrum changes, but the separation between the
two remains very nearly constant, suggesting a beat interpretation
(Strohmayer et al 1996; Miller, Lamb \& Psaltis 1998), although this
is not always so in observation or theory (van der Klis et al 1996;
Titarchuk, Lapidus \& Muslimov 1998).  A third, nearly coherent
oscillation is also seen sometimes during type I thermonuclear bursts,
near frequencies either one or two times the separation frequency
between the two kilohertz peaks, and suggests that the neutron star
spin frequency is observed directly.


In this Letter we present our findings from \xte\ observations of the
atoll source \fouru\ (Oosterbroek et al 1991), on 1997 July 19--30
when it exhibited both two kilohertz QPO peaks and bursts with
oscillations.  Preliminary results from a part of the data have
already been presented by Strohmayer, Swank \& Zhang (1998).  Strong
kilohertz oscillations from 625--925~Hz were easily detected in
several portions of the persistent emission of the source.  A second,
much broader kilohertz peak was also detected at higher frequencies
using a sensitivity-enhancing averaging technique advanced by
M\'{e}ndez et al (1998a).  We also detected QPO and noise features in
the 1--100~Hz frequency range, which changed during the observation.
Six type I bursts were seen, of which five exhibited sharp QPOs near
330~Hz.

\section{Observations}
\begin{figure}[tbp]   
\centerline{\epsfxsize=5.25in\epsfbox{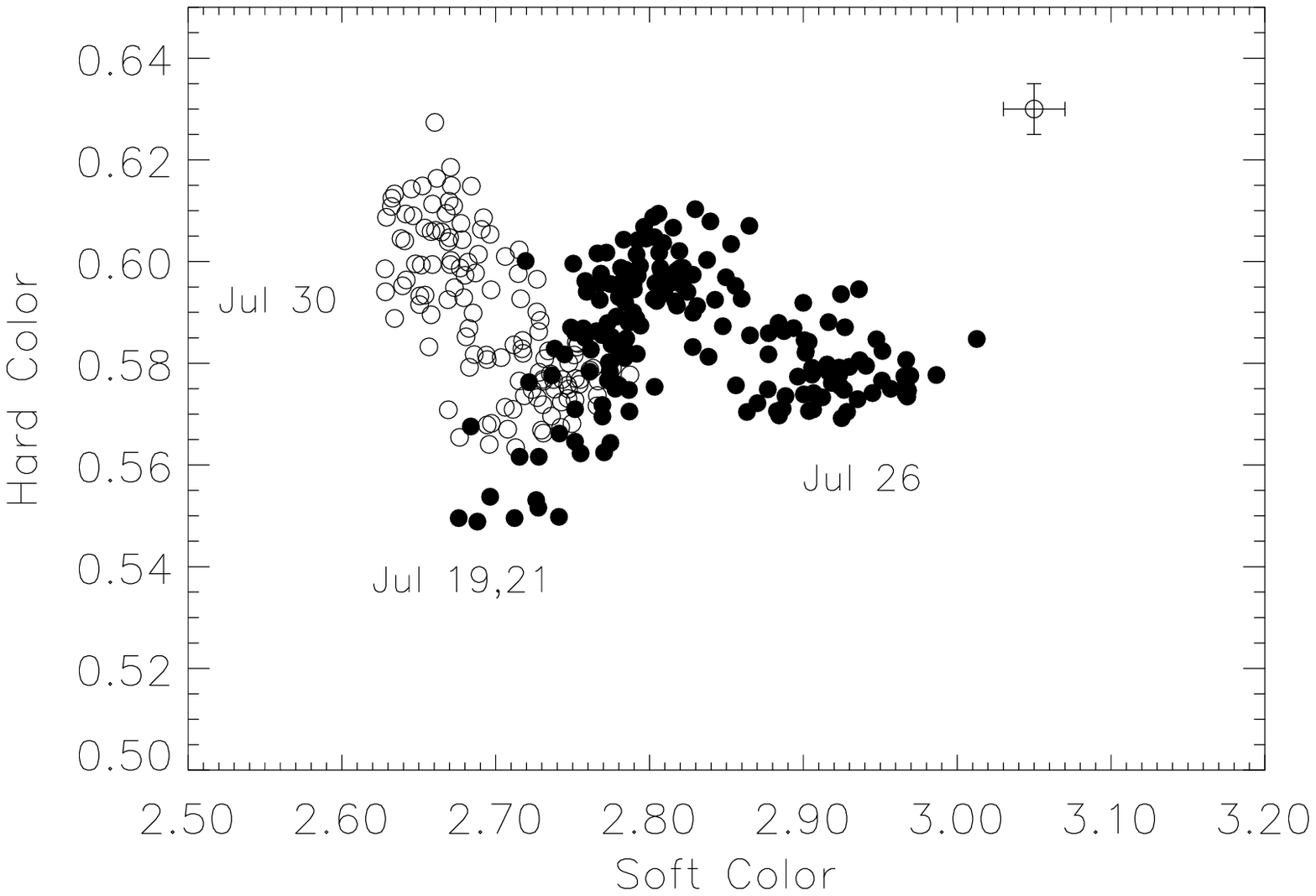}} 
\caption{ 
X-ray color-color diagram of 4U 1702-429. Each point
represents a 256 s average over the 16 s ratios, excluding bursts,
when all 5 detectors were on. Intervals in which kilohertz QPO were
observed are represented by open circles.  The filled circles
represent intervals in which no kilohertz QPO is detected.  The ``Hard
color'' is the ratio of counts in the 9.43--16.0~keV and 6.52--9.43~keV
bands and the ``Soft color'' is the ratio in the 3.63--6.52~keV and
1.86--3.63~keV bands.\label{Fcolor}}
\end{figure}

Pointed observations of \fouru\ by \xte\ Proportional Counter Array
(PCA) occurred between 1997~July 19, 21, 26, and 30, for a total
exposure of approximately 101~ks.  In addition to the ``Standard''
modes, data were also collected in an ``event'' mode with 125
microsecond timing and spectral information for individual X-ray
events.  Burst ``trigger'' and ``catcher'' modes were also used.  The
average source intensity changed significantly over the course of the
observation; average background subtracted rates (2.5-16 keV) were 670
s$^{-1}$ July 19 and July 21, 990 s$^{-1}$ July 26, and 859 s$^{-1}$
July 30.  We used the background models based on the ``Very Large
Event'' (VLE) rate and the activation rate during non-source sky
observations (model version 1).  The color-color diagram is shown in
Figure~\ref{Fcolor}.


The color differences reflect spectral differences which are
significant, although not large. For the persistent flux, the
Standard~2 spectral data were gathered for intervals with similar
temporal behavior, using only data up to 30~keV in the top layer of
detector anodes of all proportional counters.  A systematic error of
1.5\% was added (this error was needed to fit contemporaneous Crab
data).  Acceptable fits were obtained with a cut off power law
spectrum $E^{-\gamma} \times \exp(-E/kT)$. For July 26 (10.3--13.8 UT)
the photon index $\gamma=0.35\pm 0.03$ and $kT=3.53\pm 0.04$~keV,
while for July 30 (8.5--12.0 UT) $\gamma$=0.80 $\pm$ 0.03, and
$kT=4.56\pm 0.07$. Plausibly, the steeper power law and higher $kT$
correspond to upper left points in Figure~\ref{Fcolor}, the flatter
power law and lower $kT$ to the rightmost points in the figure. The
flux ranged from 1.7(1.3)$\times 10^{-9}$ erg cm$^{-2}$ s$^{-1}$ on
July 19 and 21 to 2.6(2.0)$\times 10^{-9}$ erg cm$^{-2}$ s$^{-1}$ on
July 26, for 1.6--30 (2--10) keV.  These values overlap those reported
for previous observations (Christian \& Swank 1997; Oosterbroek et al
1991).

\section{Kilohertz Variability \label{Skhz}}
Event data for channels 0--79 (0.03--29.1~keV) were used to construct
power spectra in 8~s segments, which were then averaged in groups of
16 (i.e., averaged over 128~s) to increase signal to noise.  Power
spectra on July~19 have a faint but significant QPO feature near
890--910~Hz with a full width at half maximum (FWHM) of 32~Hz and RMS
amplitude of 6.2\%.  The frequency changes significantly during the
observation.  A narrow but lower frequency QPO appears in the July~30
observations, during which the frequency of the QPO wanders
continuously between 625--825~Hz, with a mean frequency of 722~Hz.
The presence and absence of these high frequency QPOs is indicated in
Figure~\ref{Fcolor}.

\begin{figure}[tbp]   
\centerline{\epsfxsize=5.25in\epsfbox{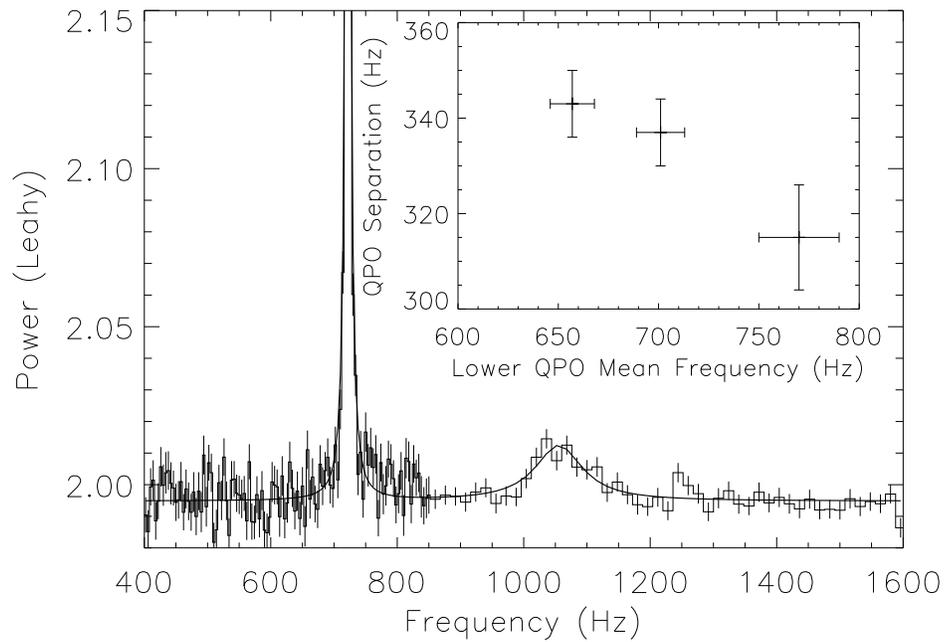}} 
\caption{ 
Average 2--25~keV PCA power spectrum for observations on
July~30, after shifting the centroid of the lower peak to 722~Hz.  The
lower half of the spectrum has 4~Hz frequency bins, while the upper
half is rebinned to 16~Hz bins for display purposes to enhance the
broad faint 1050~Hz peak.  The inset shows the fitted QPO separation
as a function of the mean lower QPO frequency (see text).  Horizontal
error bars represent the standard deviation of frequencies in the
chosen band.\label{Fshiftadd}}
\end{figure}

A heavily rebinned spectrum of July~30 indicated a much fainter and
broader peak at about 1000~Hz was present, which tracked the lower
peak.  To more precisely determine the parameters of the upper and
lower peak, the ``shift and average'' technique was employed to align
the lower peak.  This technique was first applied to find a weaker
upper QPO peak in 4U~1608--52 (M\'{e}ndez et al 1998a), and is optimal
when the peak separation remains constant or nearly so.  The centroid
of the lower strong and narrow peak was fitted in each 128~s spectrum
individually.  The spectra were then shifted so that the centroid
frequency became the mean fitted frequency (722~Hz).  Finally the
spectra were averaged, producing a strong single peak at 722~Hz.
(Figure~\ref{Fshiftadd}).  The same technique was applied to data from
July~19, except that all peaks were aligned to the average frequency
of 902~Hz.

The power spectrum of July~30 clearly shows a second broader peak at
approximately 1050~Hz (the average centroid frequency of the upper
peak).  We fitted the average July~30 spectrum with a model composed
of a constant (representing Poisson noise level) and two Lorentzian
peaks, where the upper peak centroid frequency was parameterized in
terms of an offset $\Delta\nu$ from the lower peak.  The best-fit
parameters of the model are in Table~\ref{Tqpo}, and show the upper
peak was detected at a very high significance on July~30 at a
separation of $333\pm 5$~Hz from the lower peak.  On July~19 when the
narrow QPO was present at $\sim 900$~Hz there was no evidence for a
second peak; the 95\% confidence limits for the RMS amplitude of QPOs
having a FWHM 10 and 100~Hz are 1.7\% and 2.9\%, respectively.  No QPO
peaks were present on July~21 or July~26 individually.  Combined, the
95\% confidence upper limits for both days are 1.2\% and 2.0\% for
QPOs with FWHMs of 10 and 100~Hz, respectively.  


Both QPOs are similar to those observed from other kilohertz QPO
sources.  In cases where the high frequency QPOs were detected, the
fractional RMS amplitude increases with energy.  On July~30, the RMS
fraction of the lower peak increased monotonically from 5.5\% to
10.7\% to 13.1\% in the channel ranges 0 to 17 (0.03--6.2~keV), 18 to
26 (6.5--9.4~keV) and 27 to 79 (9.8--29.1~keV); the amplitude was
about 30\% less on July~19.  The amplitude of the upper peak on
July~30, the only day it was detected, increased from 3.5\% to 6.1\%
to 12.5\% in the same energy bands.  

In both Sco X-1 and 4U~1608--52, the separation between upper and
lower QPO frequencies is not a constant, but rather decreases with
either count rate or the lower QPO frequency (van der Klis et al 1997;
M\'{e}ndez et al 1998b).  On July~30, when the upper peak is detected,
the centroid frequency of the lower peak varies from 625--825~Hz.  We
attempted to determine whether the frequency separation also varies by
grouping the individual 128~s power spectra in ranges according to the
lower QPO centroid frequency before shifting: $<$675~Hz, 675--725~Hz,
and $>$725~Hz.  Above 725~Hz the upper peak becomes weaker, and
degraded statistics prevent further divisions. The groups were shifted
and averaged, and fitted by the same model as in Table~\ref{Tqpo}.
The inset of Figure~\ref{Fshiftadd} shows the fitted frequency
separation as a function of the lower QPO frequency.  The three
spectra were also fitted to a model where the separation was held
fixed.  An $F$-test indicates that the hypothesis of constant
separation can be rejected at only a 1.5$\sigma$ confidence level.
Nevertheless, the general trend in Figure~\ref{Fshiftadd} is similar
to that seen in both 4U~1608--52 and Sco X-1.

\section{Lower Frequency Variability \label{Sbroad}}

\begin{figure}[tbp]   
\centerline{\epsfxsize=5.25in\epsfbox{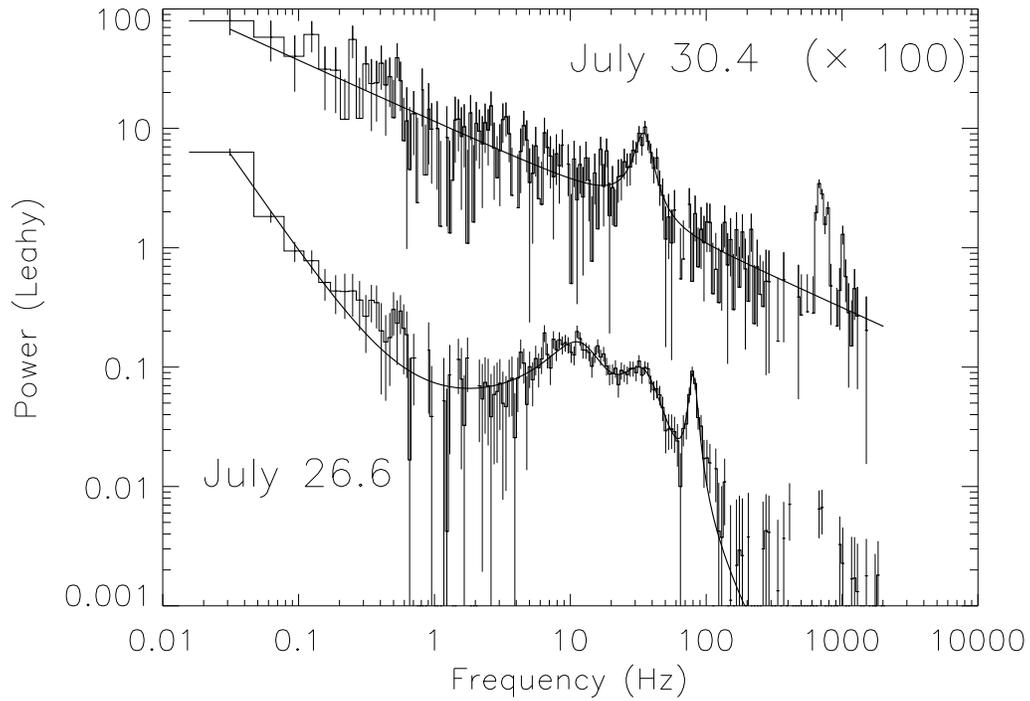}} 
\caption{ 
Broadband power spectrum of \fouru\ starting on July~26.6
(bottom curve) and July~30.4 (top curve) in the channel range 18 to 79
(6--26.3~keV), after subtracting the Poisson noise level.  The upper
curve has been offset by two decades for clarity.  The continuous
curves indicate the best fitting power law plus Lorentzian model for
each power spectrum.  A $\sim$800~Hz QPO peak, unaligned in this
representation, appears on July~30 but did not enter into the
fit.\label{Fbroadband}}
\end{figure}

\fouru\ also exhibits transient variability at frequencies below
100~Hz; at several points during the observation both QPO and broad
band noise features were observed.  Figure~\ref{Fbroadband} shows a
representative sample of average unshifted\footnote{The linear
shifting process used to align the 625--825~Hz and 900~Hz QPOs
corrupts the power spectrum at lower frequencies.} 64~s power spectra.
Between July~26.6--26.9, when no kilohertz QPOs were present, the
source had a very strong 80~Hz QPO and a lower frequency band limited
complex of variability between 5--60~Hz.  We fitted the power spectrum
with a model consisting of a constant, a power law, and up to three
Lorentzian functions.  The best fit QPO parameters for the
observations are given in Table~\ref{Tqpo}, and the representative
models are plotted in Figure~\ref{Fbroadband}.  When the two kilohertz
QPOs returned on July~30, only a single 35~Hz QPO peak was seen
(Figure~\ref{Fbroadband}, upper curve).  On both days, the continuum
extended to about 1~mHz, where a turnover was evident.  The noise
continuum changed over the observation, with power law indices of
$-0.79\pm 0.06$, $-1.68\pm 0.09$, $-1.63\pm0.08$, and $-0.517\pm
0.035$ on July~19, 21, 26, 30, respectively.  The integrated low
frequency noise ($5\times 10^{-4} - 3\times 10^{-2}$~Hz) was 1.8, 3.4,
4.0 and 1.7\% for the same days.

Although there does not appear to be a direct harmonic relationship
between any of the QPO frequencies below 100~Hz, the centroid of the
35~Hz peak on July~30 is correlated with the centroid of the
simultaneously-present kilohertz QPOs.  We found that as the mean
kilohertz QPO frequency increased in each band from 657~Hz to 702~Hz
to 769~Hz, the low frequency QPO centroid increased from $32.9\pm
0.6$~Hz to $34.8\pm 0.6$~Hz to $40.1\pm 0.8$~Hz, thus establishing a
ratio of $\sim 19.5$ between the two.  This linkage between the
kilohertz and lower frequency QPOs is similar to that seen previously
(e.g. Sco X-1, van der Klis et al 1996; 4U~1728$-$34, Ford \& van der
Klis 1998).

\section{Burst Oscillations}
\begin{figure}[tbp]   
\centerline{\epsfxsize=4.5in\epsfbox{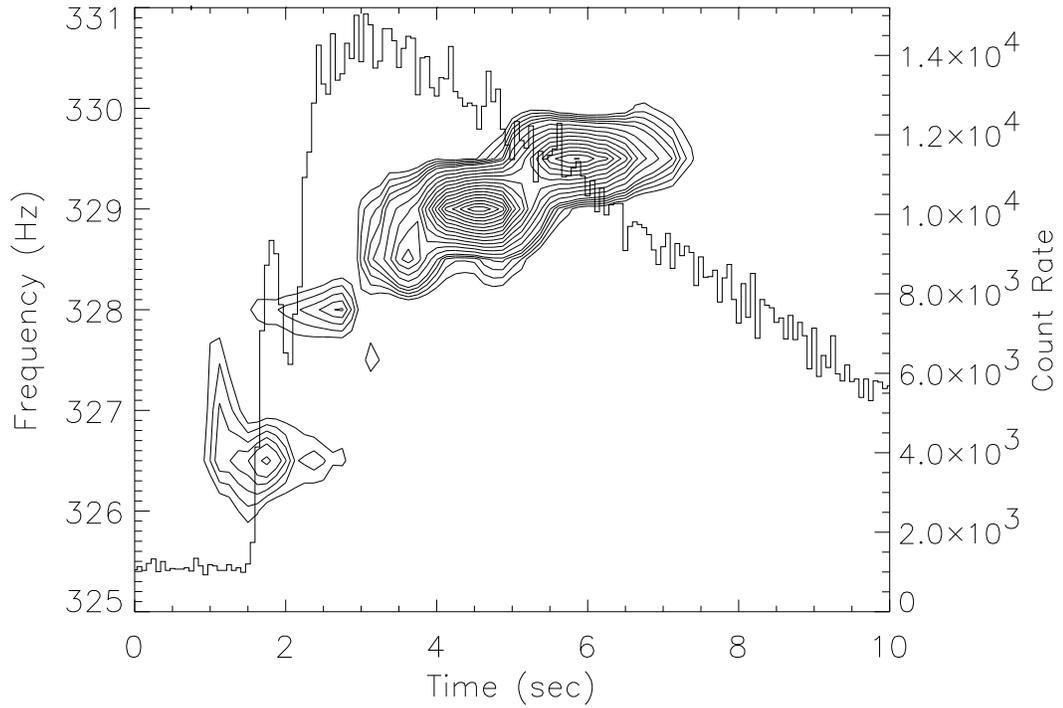}} 
\caption{ 
Dynamic power spectrum showing $\approx 330$ Hz pulsations
during a burst from \fouru. The contours denote loci of constant
Fourier power as a function of frequency (left ordinate) and time. The
individual power spectra were computed from 2 s intervals sampled at
1/8192~s time resolution, with a new interval beginning every
0.125~s. The PCA countrate versus time is also shown (right
ordinate). The power spectral contours were aligned on the center of
each 2 s interval.\label{Fburst}}
\end{figure}

A total of six type I (thermonuclear) X-ray bursts occurred during the
observations. The peak PCA count rate (2--90 keV) during the bursts
ranged from a low of 14,000~ct~\psec\ to a maximum of 26,000~ct~\psec.
To search for pulsations we used the 2--90~keV burst mode high time
resolution data to compute power spectra from consecutive 2~s
intervals beginning near the peak of each burst.  We detected
pulsations at $\approx 330$~Hz in five of the six bursts.  Pulsation
amplitudes in the 2--24~keV band ranged from a few percent to as high
as 18\% (RMS).  The pulsation frequency increases during all the
bursts, eventually reaching an upper, limiting frequency in the
decaying tail of the burst. This behavior is nearly identical to the
frequency evolution of pulsations observed in bursts from 4U~1728--34
and other sources with oscillations during bursts (see Strohmayer et
al. 1998; and Strohmayer, Swank \& Zhang 1998). To illustrate this
behavior we computed dynamic power spectra for each burst using 2~s
intervals with a new interval beginning every 0.125~s.
Figure~\ref{Fburst} shows such a dynamic power spectrum computed for
the burst observed on 1997~July 26 at 14:04:19~UT, with the PCA
countrate overlaid. The run of frequency during this burst of
relatively low peak luminosity was similar to that of the high
luminosity burst shown in Strohmayer, Swank, \& Zhang (1998).  We will
present a more comprehensive timing and spectral analysis of the
bursts in a subsequent publication.

\section{Discussion and Conclusions}

Atoll burst sources range in estimated luminosity from a hundredth to
perhaps a third of the Eddington limit for neutron stars.
Kilohertz-range QPOs have been seen in sources throughout this
luminosity range. \fouru\ is an atoll burst source with a luminosity,
assuming a distance of 7~kpc (Oosterbroek et al 1991), of about
$1\times 10^{37}$~erg~\psec\, which is similar to that of 4U~1728$-$34
and 4U~1705$-$44. We found that \fouru, like 4U~1728$-$34 (Strohmayer
et al. 1996) also exhibits two kilohertz QPOs with a separation of
about 330~Hz, to be compared to 363~Hz for 4U~1728$-$34.  Furthermore,
five bursts exhibited flux oscillations within a few Hertz of 330~Hz.
This makes \fouru\ only the second source in which the dominant burst
power is at the frequency of the difference between the two higher
frequencies seen in the persistent flux.  The lower QPO peak has been
interpreted as a beat between the upper peak and the neutron star spin
frequency (Miller, Lamb \& Psaltis 1998). The separation between the
two kilohertz peaks is approximately constant for several sources, and
is thought to be the neutron star spin frequency.  Our results here
are consistent with a constant separation, although a small variation
is suggested.


On July~30, when both peaks are observed, the highest frequency of the
upper QPO is inferred to be at least 1156~Hz, assuming the peak
separation is 333~Hz.  Given that the lower peak on July~30 is quite
narrow compared to the upper one, it seems likely that the single peak
that appears on July~19 is also a ``lower'' peak based on its width.
If that is so, then we might speculate that the maximum upper
frequency, although not detected, is as high as $\sim 1230$~Hz.  In
either case, the range of 1100--1200~Hz is consistent with the maximum
frequency observed in other kilohertz QPO sources. The small changes
in burst frequency during the burst are similar to those of
4U~1728$-$34 (Strohmayer et al. 1996) and the oscillations appear
equally as consistent with rotation of a single spot on the neutron
star.

The \xte\ All-Sky Monitor record of \fouru\ shows that the source
intensity range is about 25--100~mCrab.  During the week spanning the
observations, the range was about 70--90~mCrab, yet the source
exhibited markedly varied behavior.  The atoll sources were defined by
Hasinger and van der Klis (1989) in terms of both crude spectral
behavior (color-color diagrams) and the continuum shape of the power
spectra, based on {\it EXOSAT} data.  The hard and soft colors trace a
curve reminiscent of an atoll, comprised of an ``island'' and a
crescent shaped ``banana'' branch.  In the power spectra low frequency
noise is prevalent on the banana branches, and flat high frequency
noise identifies the island state. In \xte\ data a pattern has emerged
in which the kilohertz QPO appear to be present in the lower banana
and at least part of the island state (e.g., 4U~1608$-$52, M\'{e}ndez
et al 1998a; Aql X-1, Cui et al 1998; 4U~1735$-$44, Wijnands et al
1998; 4U~1820$-$30, Zhang et al. 1998).  Even with the limited
excursions of our observations, the color-color diagram suggests that
\fouru\ behaves similarly.  If we follow the behavior of the other
sources, then we would predict that the kilohertz QPOs would fade out
during the flares and possibly at minimum flux.

These observations may not have quite reached the island state, which
is described as having band limited white noise with a cut-off near
the low-frequency QPO.  The continuum portion of the power spectrum of
July~30, while shallower than July~26, was not flat.  On the other
hand, the power spectrum of July~26 which has a steep low frequency
component, and is thus presumably in an intermediate banana state,
also has unaccounted-for QPO-like structures.  We should note that the
color-color diagram suggests the banana branch may not be unique.

A strong 35~Hz feature accompanies the kilohertz QPOs, and while the
lower kilohertz QPO does not have large excursions, the two do appear
to track each other.  This behavior might be expected if the 35~Hz QPO
represented Lense-Thirring precession of a warped inner accretion disk
under the influence of a spinning compact body (Stella \& Vietri
1998).  The theory predicts oscillations in the tens of hertz range,
depending on the ratio of the moment of inertia ($I=I_{45}
10^{45}$~g~cm$^2$) to the mass of the neutron star $M$.  Assuming the
neutron star spin frequency is 330~Hz, the Kepler frequency at the
inner edge of the accretion disk is 330~Hz higher than the lower peak,
and the ratio $I_{45}M_{\sun}/M$ is a free parameter, we fit to the
functional form of Stella \& Vietri (1998), and find a ratio of
$2.3\pm 0.1$.  Although this value is too high for most neutron star
equations of state, which approach unity (Markovi\'{c} \& Lamb 1998),
it is closer than the reported ratios of 4--5 for most other QPO
sources with such measurements (e.g., 4U~1735$-$44, Wijnands et al
1998; 4U~1728$-$34, Ford \& van der Klis 1998).


The catalog of low frequency QPOs is indeed complex.  A 35--40~Hz QPO
accompanies the kilohertz QPOs but with varying width.  The power
spectrum of July~26 also shows a $\sim$30~Hz QPO, as well as
$\sim$10~Hz and 80~Hz components, while on July~21 a peak at
$\sim$30~Hz is not present.  Such QPOs have been detected in other
atoll sources and may be comparable to the horizontal branch
oscillations (HBO) of Z sources.  However, the 80~Hz frequency
observed here is one of the highest so far seen below 100~Hz in a
neutron star system.  The power spectrum of GX 13+1, an atoll source
several times brighter than \fouru, has interesting similarities to
the power spectra we have seen for \fouru\ when the kilohertz QPO were
not seen.  Homan et al. 1998 found a 57--61~Hz QPO when the source was
in an upper part of a banana branch in the color-color diagram, along
with very low frequency noise described by a power law of index
$-1.3$, and a peaked noise component. The power spectrum of \fouru\ in
the highest part of its banana branch contains similar components.
Homan et al suggest the peaked noise and the QPO in GX~13+1 correspond
to the low frequency noise (LFN) and HBO in Z sources and to the high
frequency noise and low frequency QPO in atoll sources. It is
interesting that \fouru\ exhibits the atoll low frequency QPO
(although not the noise) when the kilohertz QPO is present, and is
similar to GX~13+1 when it is not present.


\newpage

\newpage

\begin{table}
\caption{Temporal Properties of \fouru\label{Tqpo}}
\begin{tabular}{lrlcc}
\hline
Date    & \multispan{2}{Frequency (Hz)} & RMS (\%) & FWHM (Hz) \\
\hline
Jul 19.37--19.86 & $\nu$&$= 902$\tablenotemark{a} & $6.0\pm 0.1$ & $6.1\pm 0.3$\\
        & $\nu$&$= 39.0\pm 3.4$ & $5.3\pm 0.8$ & $27\pm 10$\\
        
\hline
Jul 21.01-21.17  & $\nu$&$= 85.6\pm 3.7$ & $9.6\pm 0.8$ & $51.5\pm 11.6$\\
        & $\nu$&$=12.0\pm 1.0$  & $7.7\pm 0.6$ & $15.7\pm 0.9$\\
        
\hline
Jul 26.60--26.93  & $\nu$&$= 80.1\pm 0.5$ & $6.0\pm 0.3$ & $12.0\pm 1.7$\\
        & $\nu$&$= 32.5\pm 1.2$ & $9.1\pm 1.1$ & $26.2\pm 3.6$\\
        & $\nu$&$= 10.8\pm 0.4$ & $7.9\pm 0.5$ & $12.4\pm 1.6$\\
\hline
Jul 30.46--30.72 & $\nu$&$= 722$\tablenotemark{a} & $7.78\pm 0.08$ & $7.22\pm 0.18$\\
     & $\Delta\nu$&$= 333\pm 5$\tablenotemark{b}& $5.33\pm 0.40$ & $91\pm 17$\\
        & $\nu$&$= 34.7\pm 0.6$ & $6.3\pm 0.4$ & $12.6\pm 2.0$ \\
\hline
\end{tabular}
\tablenotetext{a}{the mean fitted QPO frequency before shifting (see text).}
\tablenotetext{b}{the frequency {\it difference} from the lower peak.}
\end{table}

\end{document}